\documentclass[showpacs,amsmath,
twocolumn,
aps,prl]{revtex4}
\usepackage{graphicx}
\usepackage{dcolumn}
\usepackage[colorlinks=true, pdfstartview=FitV, linkcolor=blue, 
            citecolor=blue, urlcolor=blue]{hyperref}

\begin{document}

\title{Feedback cooling of a cantilever's fundamental mode below 5 mK}
\author{M. Poggio$^{1,2}$, C. L. Degen$^1$, H. J.  Mamin$^1$ and D.
  Rugar$^1$} \affiliation{$^1$IBM Research Division, Almaden Research
  Center, 650 Harry Rd., San Jose CA, 95120 \\ $^2$Center for Probing
  the Nanoscale, Stanford University, 476 Lomita Hall, Stanford CA,
  94305} \date{\today}

\begin{abstract}
  We cool the fundamental mechanical mode of an ultrasoft silicon
  cantilever from a base temperature of 2.2 K to $2.9 \pm 0.3$ mK
  using active optomechanical feedback.  The lowest observed mode
  temperature is consistent with limits determined by the properties
  of the cantilever and by the measurement noise.  For high feedback
  gain, the driven cantilever motion is found to suppress or
  ``squash'' the optical interferometer intensity noise below the shot
  noise level.
\end{abstract}

\pacs{85.85+j, 42.50.Lc, 45.80.+r, 46.40.-f}

\maketitle

Feedback control of mechanical systems is a well-established
engineering discipline which finds applications in diverse areas of
physics, from the stabilization of large cavity mirrors used in
gravitational wave detectors \cite{Abbott:2004} to the control of tiny
cantilevers in atomic force microscopy
\cite{Albrecht:1990,Mertz:1993,Garbini:1996,Bruland:1998,Weld:2006}.
Recently, the prospect of cooling a mechanical resonator to its
quantum ground state has spurred renewed interest in the damping of
oscillators through both active feedback \cite{Kleckner:2006} and
passive back-action effects \cite{Naik:2006,Gigan:2006,Arcizet:2006}.
Motivated by the ability to make ever smaller mechanical devices and
ever more sensitive detectors of motion, researchers are pushing into
a regime in which collective vibrational motion should be quantized
\cite{Schwab:2005}.  In combination with conventional cryogenic
techniques, the cooling of a single mechanical mode using feedback may
provide an important step towards achieving the quantum limit in a
mechanical system.  Here we demonstrate the feedback cooling of an
ultrasoft silicon cantilever to below 5 mK starting from a base
temperature as high as 4.2 K.  Starting from this temperature, the
vibrational mode of the oscillator is cooled near the level of the
measurement noise, which sets a fundamental limit on the cooling
capacity of feedback damping.  In the future, minimizing such noise
may be key to achieving single-digit mode occupation numbers.

\begin{figure}[t]\includegraphics{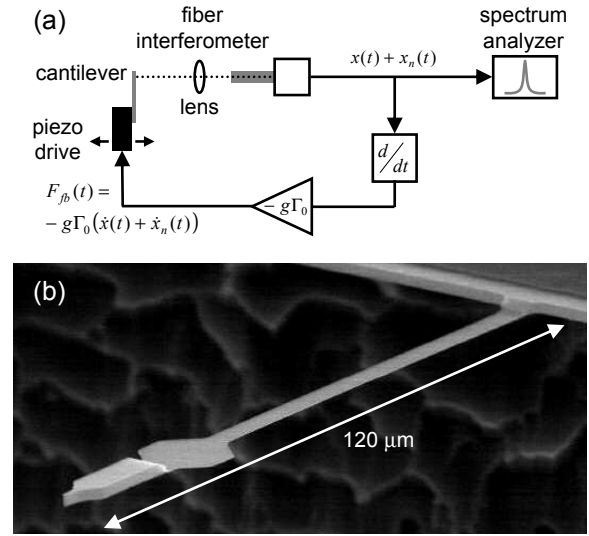}
\caption{\label{fig1}
  (a) Schematic diagram of the experimental setup and (b) scanning
  electron micrograph of a representative Si cantilever.
}\end{figure}

We study the fundamental mechanical mode of two $120 \times 3 \times
0.1$-$\mu$m single-crystal Si cantilevers of the type shown in
Fig.~\ref{fig1}(b).  The ends of the levers are designed with a $2
\times 15$-$\mu$m mass which serves to suppress the motion of flexural
modes above the fundamental \cite{Chui:2003}.  Cantilevers 1 and 2
have resonant frequencies of 3.9 and 2.6 kHz respectively due to the
difference in mass of the samples which have been glued to their ends.
The sample on cantilever 1 is a 0.1-$\mu$m$^3$ particle of SmCo while
the sample on cantilever 2 is a 50-$\mu$m$^3$ particle of CaF$_2$
crystal.  Both samples are not related to the work presented here
aside from the added mass which they contribute.  The oscillators'
spring constants are both determined to be $k = 86$ $\mu$N/m through
measurements of their thermal noise spectra at several different base
temperatures.  Each cantilever is mounted in a vacuum chamber
(pressure $< 1 \times 10^{-6}$ torr) at the bottom of a dilution
refrigerator which has been isolated from environmental vibrations.
The motion of the lever is detected using laser light focused onto a
10-$\mu$m-wide paddle near the mass-loaded end and reflected back into
an optical fiber interferometer \cite{Rugar:1989}.  100 nW of light
are incident on the lever from a temperature-tuned 1550-nm distributed
feedback laser diode \cite{Bruland:1999}.  The interferometric
cantilever position signal is sent through a differentiator circuit
and a variable electronic gain stage back to a piezoelectric element
which is mechanically coupled to the cantilever, as shown
schematically in Fig.~\ref{fig1}(a).  The overall bandwidth of the
feedback is controlled by bandpass filters.  For negative gain, this
feedback loop has the effect of producing a damping force on the
cantilever proportional to the velocity of its oscillatory motion.

For frequencies in the vicinity of the fundamental mode resonance, the
motion of a cantilever is well approximated by

\begin{equation}
\label{eq1}
m \ddot{x} + \Gamma_0 \dot{x} + k x = F_{th} - g \Gamma_0 \left ( \dot{x} + \dot{x}_n \right ),
\end{equation}

\noindent
where $x(t)$ is the displacement of the oscillator as a function of
time, $\Gamma_0$ is its intrinsic dissipation, $k = m {\omega_0}^2$ is
its spring constant, $m$ is the oscillator's effective mass,
$\omega_0$ is its angular resonance frequency, $F_{th}(t)$ is the
random thermal Langevin force, and $x_n(t)$ is the measurement noise
on the displacement signal.  The dissipation can be written in terms
of $m$, $\omega_0$, and the oscillator's intrinsic quality factor
$Q_0$ according to $\Gamma_0 = m \omega_0 / Q_0$.  

Given the equation of motion in (\ref{eq1}) and considering frequency
components of the form $\hat{F}_{th}(\omega) e^{i \omega t}$ and
$\hat{x}_n(\omega) e^{i \omega t}$, the frequency response of the
oscillator is

\begin{equation}
\label{eq1.5}
\hat{x}(\omega) = \frac{\frac{1}{m} \hat{F}_{th}(\omega) - i g \frac{\omega_0 \omega}{Q_0} \hat{x}_n(\omega)}{\left ( {\omega_0}^2 - \omega^2 \right ) + i (1 + g) \frac{\omega_0 \omega}{Q_0}}.
\end{equation}

\noindent
For random excitations where $F_{th}(t)$ and $x_n(t)$ are
uncorrelated, we can then determine the spectral density of both the
oscillator's {\it actual} displacement $x$,

\begin{eqnarray}
\label{eq3}
S_{x}(\omega) = & \left [ \frac{1 / m^2}{\left ( {\omega_0}^2 - \omega^2 \right )^2 + \left ( 1 + g \right )^2 {\omega_0}^2 \omega^2 / {Q_0}^2} \right ] S_{F_{th}} \nonumber & \\
& + \left [ \frac{g^2 {\omega_0}^2 \omega^2 / {Q_0}^2}{\left ( {\omega_0}^2 - \omega^2 \right )^2 + \left ( 1 + g \right )^2 {\omega_0}^2 \omega^2 / {Q_0}^2} \right ] S_{x_n}, &
\end{eqnarray}

\noindent
and of its {\it measured} displacement $x + x_n$,

\begin{eqnarray}
\label{eq2}
S_{x + x_n}(\omega) = & \left [ \frac{1 / m^2}{\left ( {\omega_0}^2 - \omega^2 \right )^2 + \left ( 1 + g \right )^2 {\omega_0}^2 \omega^2 / {Q_0}^2} \right ] S_{F_{th}} \nonumber & \\
& + \left [ \frac{\left ( {\omega_0}^2 - \omega^2 \right )^2 + {\omega_0}^2 \omega^2 / {Q_0}^2}{\left ( {\omega_0}^2 - \omega^2 \right )^2 + \left ( 1 + g \right )^2 {\omega_0}^2 \omega^2 / {Q_0}^2} \right ] S_{x_n}. &
\end{eqnarray}

\noindent
Here $S_{x_n}$ is the spectral density of the measurement noise $x_n$
and $S_{F_{th}}$ is the white spectral density of the thermal noise
force $F_{th}$.  According to the fluctuation dissipation theorem, the
noise force $S_{F_{th}}$ depends on the cantilever dissipation and is
given by $S_{F_{th}} = 4 \Gamma_0 k_B T$, where we are using a
single-sided convention for the spectral density.

We define the mode temperature of the cantilever according to the
equipartition theorem as $T_{\text{mode}} = k \left \langle x^2 \right
\rangle / k_B$ and calculate $\left \langle x^2 \right \rangle$
according to $\left \langle x^2 \right \rangle = (1 / 2 \pi)
\int_0^\infty {S_x(\omega) d \omega}$.  Using (\ref{eq3}) and assuming
that $S_{x_n}$ is independent of $\omega$, we find

\begin{equation}
\label{eq4}
T_{\text{mode}} = \frac{T}{1 + g} + \frac{k \omega_0}{4 k_B Q_0}
\left ( \frac{g^2}{1 + g} \right ) S_{x_n},
\end{equation}

\noindent
where $T$ is the bath temperature and $k_B$ is the Boltzmann constant.

In the limit of small gain ($g << 1$), the effect of measurement noise
on the oscillator displacement can be ignored and the oscillator power
spectrum is obtained by simply subtracting the measurement noise floor
from the measured spectrum: $S_x(\omega) = S_{x + x_n}(\omega) -
S_{x_n}$.  The same is true for large gain as long as the noise is
well below the measured displacement power (more precisely, $ S_{x_n}
<< \frac{{Q_0}^2}{g^2 k^2} S_{F_{th}}$).  In both cases
the mode temperature is proportional to the integrated area between
the measured spectrum and the noise floor and reduces to the familiar
$T_{\text{mode}} = \frac{T}{1 + g}$ \cite{Kleckner:2006}.  Increasing
the damping gain lowers the mode temperature leading to ``feedback
cooling'' or ``cold damping'' of the oscillator.

\begin{figure}[t]\includegraphics{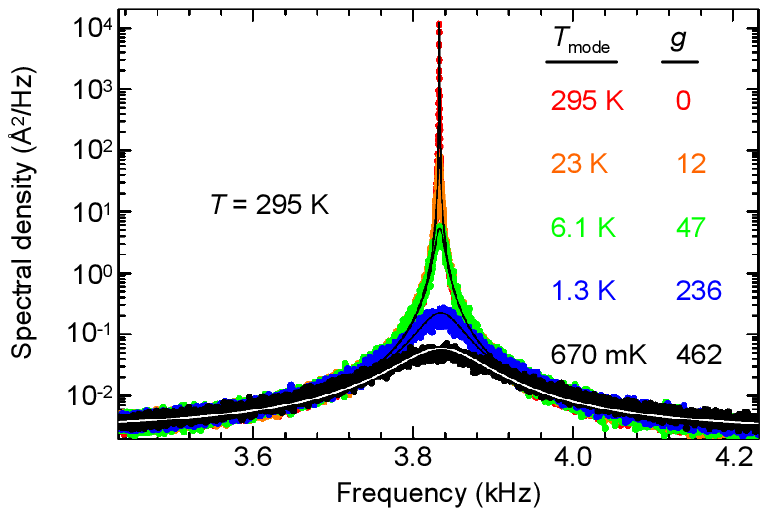}\caption{\label{fig2}
    (Color) Measured spectral density $S_{x + x_n}$ of cantilever 1
    for different feedback gains $g$ as it is cooled from a base
    temperature $T = 295$ K.  Colored data points correspond to the
    mode temperatures (with an error of $\pm 10$\%) and gains of the
    same color shown to the right.  Solid lines are fits to the data
    using (\ref{eq2}).}\end{figure}

The feedback cooling of cantilever 1 from a base temperature of 295 K
falls in this limit and is shown in Fig.~\ref{fig2}.  At this
temperature $Q_0 = 16,000$.  As the gain increases, the mode
temperature decreases down to $T_{\text{mode}} = 670 \pm 70$ mK for $g
= 462$.  Even at the highest gain, the measurement noise is well below
the observed thermal noise.  Therefore, the temperature of the
fundamental lever mode is well determined by the area between the
observed peak and the noise floor.  The mode temperatures shown in
Fig.~\ref{fig2} are equal within the error whether they are calculated
by simply integrating the area under the observed spectra or whether
the spectra are fit using (\ref{eq2}) and the extracted parameters are
substituted in (\ref{eq4}).  The fits, which are shown as solid lines
in Fig.~\ref{fig2}, involve three free parameters: $\omega_0$, $g$,
and $S_{x_n}$.

\begin{figure}[t]\includegraphics{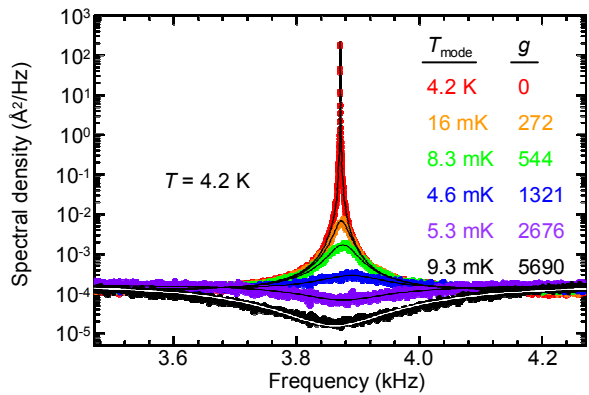}\caption{\label{fig3}
    (Color) Measured spectral density $S_{x + x_n}$ of cantilever 1
    for different feedback gains $g$ as it is cooled from a base
    temperature $T = 4.2$ K.  Colored data points correspond to the
    mode temperatures (with an error of $\pm 10$\%) and gains of the
    same color shown to the right.  Solid lines are fits to the data
    using (\ref{eq2}).}\end{figure}

When we cool cantilever 1 by feedback from a base temperature of 4.2
K, where $Q_0 = 44,200$, this approximation is no longer valid.
Starting at $g \simeq 300$, the values of $T_{\text{mode}}$ calculated
from simple integration of the spectrum above the noise floor begin to
deviate from the more accurate values given by (\ref{eq4}).
Increasing the gain further, as shown in Fig.~\ref{fig3}, pushes the
observed thermal noise spectra down to the level of the measurement
noise and beyond.  

The two spectra showing a dip below the white noise floor are clear
deviations from the low gain, low noise approximation; calculating the
mode temperature through integration would result in unphysical
negative values.  Here the feedback loop counteracts intensity
fluctuations in the light field by exciting the cantilever rather than
by damping it.  In our experiment, these intensity fluctuations are
due to the shot noise of the laser field, i.e.\ we are operating in
the limit where $S_{x_n}$ is dominated by the photon shot noise.  From
fits to the spectra, we find $\sqrt{S_{x_n}} \simeq 10^{-2}$
\AA$/\sqrt{\text{Hz}}$.  When $S_{x_n}$ is limited by shot noise, as
in our case, its suppression by feedback is known as intensity noise
``squashing'' inside an optoelectronic loop
\cite{Buchler:1999,Sheard:2005, Bushev:2006}.

\begin{figure}[t]\includegraphics{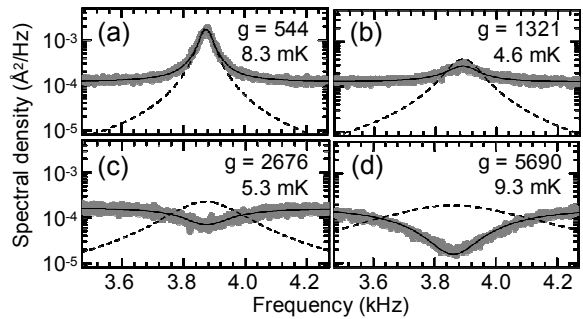}\caption{\label{fig4}
    Suppression of the thermal noise of cantilever 1 down to and below
    the measurement noise.  Gray points represent the observed
    interferometer signal $S_{x + x_n}$ with the lever at a base
    temperature of $T = 4.2$ K, solid lines are fits to this data
    using (\ref{eq2}), and dashed lines are the corresponding spectra
    of the actual cantilever motion $S_x$ as defined in (\ref{eq3}).
  }\end{figure}

In the high gain regime ($ g > 300$) of Fig.~\ref{fig3}, we must
consider the full effect of measurement noise on (\ref{eq2}) and
(\ref{eq3}) in order to extract the actual motion of the lever.  As
shown in Fig.~\ref{fig4}, the actual vibrational spectrum of lever 1
deviates from the measured spectrum as it approaches the measurement
noise.  Here, the limits of feedback cooling have been reached as
measurement noise sent back to the piezoelectric actuator acts to heat
the lever's vibrational mode.

\begin{figure}[t]\includegraphics{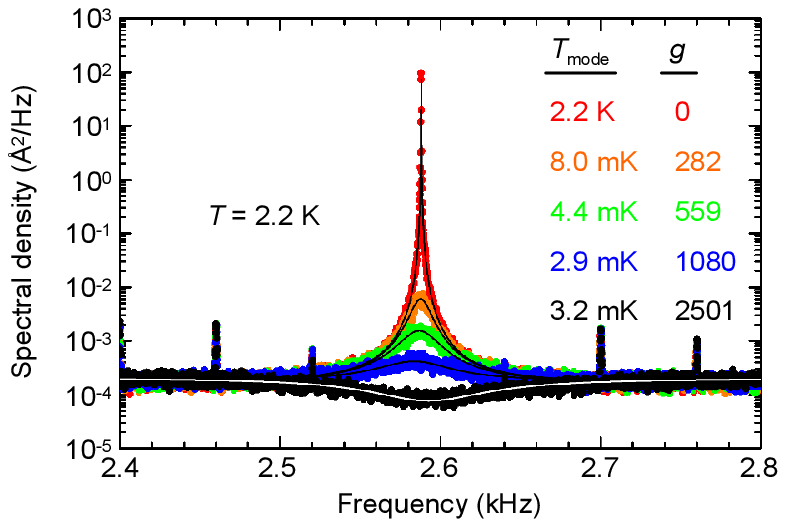}\caption{\label{fig5}
    (Color) Measured spectral density $S_{x + x_n}$ of cantilever 2
    for different feedback gains $g$ as it is cooled from a base
    temperature $T = 2.2$ K.  Colored data points correspond to the
    mode temperatures (with an error of $\pm 10$\%) and gains of the
    same color shown to the right.  Solid lines are fits to the data
    using (\ref{eq2}).}\end{figure}

We observe similar behavior from cantilever 2 starting at a lower base
temperature.  In this case, the experimental apparatus is cooled to
250 mK.  Measurement of the lever's thermal noise spectrum, however,
reveals that its base temperature reaches only 2.2 K with $Q_0 =
55,600$.  This discrepancy is due to heating of the Si cantilever
through the absorption of light from the interferometer laser.  As
shown in Fig.~\ref{fig5}, by applying feedback cooling at this base
temperature, we achieve a minimum fundamental mode temperature of $2.9
\pm 0.3$ mK before $T_{\text{mode}}$ starts increasing as a function
of $g$.

\begin{figure}[t]\includegraphics{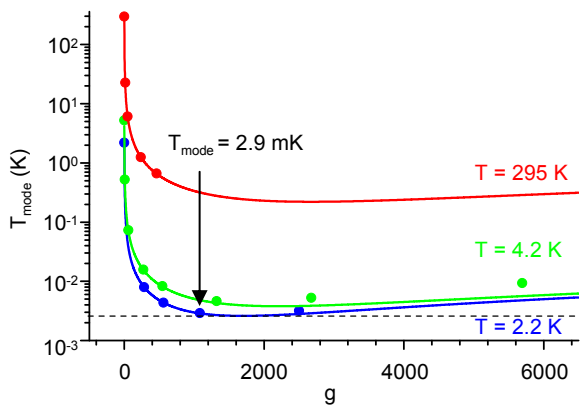}\caption{\label{fig6}
    (Color) $T_{\text{mode}}$ of our Si cantilevers as a function of
    feedback gain $g$.  Solid lines show $T_{\text{mode}}$ in
    (\ref{eq4}) for the experimental parameters and the points
    represent $T_{\text{mode}}$ extracted from three-parameter fits of
    (\ref{eq3}) to spectra including those shown in
    Figs.~\ref{fig2}-\ref{fig5}.  Lines and points are color coded to
    correspond to the colored base temperature labels.  The dashed
    line indicates $T_{\text{mode, min}} = 2.6$ mK for cantilever 2 at
    $T = 2.2$ K.  }\end{figure}

As implied by (\ref{eq4}) and shown in Fig.~\ref{fig6}, the
measurement noise floor sets a minimum achievable mode temperature for
$g >> 1$: 

\begin{equation}
\label{eq5}
T_{\text{mode, min}} = \sqrt{\frac{k \omega_0 T}{k_B Q_0}S_{x_n}}.
\end{equation}

\noindent
For cantilever 2 at $T = 2.2$ K, we calculate $T_{\text{mode, min}} =
2.6$ mK, which is close to the observed minimum temperature of $2.9
\pm 0.3$ mK.  A more complex expression could be written for
$T_{\text{mode, min}}$ if the techniques of optimal control were used
to cool the lever rather than simple velocity-proportional damping
\cite{Garbini:1996,Bruland:1998}.  For our experimental parameters,
optimal control does not provide any further reduction in
$T_{\text{mode, min}}$.  We calculate, however, that in the low noise
limit ($\sqrt{S_{x_n}} < 10^{-4}$ \AA$/\sqrt{\text{Hz}}$), it could
achieve lower mode temperatures than velocity-proportional damping.

The minimum temperature in (\ref{eq5}) immediately implies a minimum
mode occupation number $N_{\text{mode, min}} = \frac{1}{\hbar}
\sqrt{\frac{k k_B T}{\omega_0 Q_0}S_{x_n}}$.  In our case, the lowest
achieved mode occupation is $N \simeq 2.3 \times 10^4$ and
$N_{\text{mode, min}} = 2.1 \times 10^4$.  Since for a cantilever
$\frac{k}{\omega_0} \propto \frac{t^2 w}{l}$, where $l$, $w$, and $t$
are its length, width, and thickness, $N_{\text{mode, min}} \propto t
\sqrt{ \frac{w T}{l Q_0} S_{x_n}}$.  Therefore, if low occupation
numbers are to be achieved by feedback cooling, the cantilevers
employed should be long and thin, have high quality factors and the
measurement should be done at low base temperature with low
measurement noise.  The geometry of our ultra-soft cantilevers is well
suited to minimize $N_{\text{mode, min}}$.  It appears, therefore,
that the most likely way to achieve further reductions in
$N_{\text{mode, min}}$ is to reduce the measurement noise, either by
using better optical interferometry or by employing a detector of
cantilever motion with intrinsically higher resolution, such as a
single electron transistor (SET).  SETs have recently achieved
$\sqrt{S_{x_n}} \sim 10^{-5}$ \AA$/\sqrt{\text{Hz}}$
\cite{Knobel:2003,LaHaye:2004,Naik:2006}.  High-frequency doubly
clamped resonators cooled cryogenically below 50 mK have achieved
occupation numbers down around 25 without feedback
\cite{LaHaye:2004,Naik:2006}.

It is worth noting that the type of feedback cooling discussed here
may be applicable to nanoelectromechanical systems in sensing
applications.  It was shown recently that as nanomechanical resonators
shrink in size, their dynamic range decreases
\cite{Postma:2005,Kozinsky:2006}.  This effect is due to a combination
of a decrease in the onset of nonlinearity and an increase in the
thermomechanical noise with decreasing size.  A resonator's dynamic
range can at least be partially recovered through feedback cooling,
which reduces the thermal noise amplitude.

Finally, optimized feedback cooling may find use in the realization of
a type of magnetic resonance force microscopy which detects nuclear
magnetic resonance at the Larmor frequency \cite{Sidles:1992}.  Such a
scheme strongly couples the cantilever thermal noise to the nuclear
spins and has the side-effect of dramatically increasing the nuclear
spin relaxation rate.  Feedback cooling could be used both to control
this lever-induced relaxation and to dramatically reduce the
temperature of the nuclear spin system.

\begin{acknowledgments}
  We thank B. W. Chui for fabricating the cantilevers and K. W.
  Lehnert for helpful discussions.  We acknowledge support from the
  DARPA QuIST program, the NSF-funded Center for Probing the Nanoscale
  (CPN) at Stanford University, and the Swiss NSF.
\end{acknowledgments}

\end{document}